# Bridging Two Dimensions: Luminescent Sensors at the Intersection of Temperature and Pressure

Lukasz Marciniak[1,*], Maja Szymczak[1], Przemysław Woźny[2], Marcin Runowski[2*]

[1] Institute of Low Temperature and Structure Research, Polish Academy of Sciences,

Okólna 2, 50-422 Wrocław, Poland

[2] Adam Mickiewicz University, Faculty of Chemistry, ul. Uniwersytetu Poznańskiego 8, 61-614 Poznań, Poland

*corresponding authors: l.marciniak@intibs.pl, runowski@amu.edu.pl

**Table of content**

**Abstract**

**Luminescence thermometry and manometry are exponentially growing areas dealing with the optical detection of temperature and pressure, respectively, being appealing alternatives for conventional thermometers and manometers. The main benefit of luminescent thermometers and manometers is a possibility of remote temperature and/or pressure monitoring, in contrast to conventional gauges. Moreover, the use of luminescent nanoparticles as temperature/pressure sensors allow detection in micron- and nano-sized areas, previously inaccessible for conventional gauges. Therefore, the combination of both functionalities in a single material is highly appealing, as has been shown in a growing number of reports in the last years. Moreover, the bifunctional pressure and temperature sensors, operating with multiple independent spectroscopic parameters, allow simultaneous and distinct pressure and temperature readouts. However, the development of such truly bifunctional and reliable sensors is very challenging and rarely reported. This review summarizes the current status in the field, focusing on the sensing strategy and the selection of optically active sensor materials appropriate for a given application, including their sensitivity, spectral range of interest and pressure/temperature (in)dependence.**

## 1. Introduction

Among the wide-ranging applications of luminescent materials, such as light-emitting diodes[1–4], solar concentrators[5,6], light converters[7–10], night vision devices[11–14], plant cultivation[15,16], bio-labeling[17,18], *etc*.,-one of the most rapidly advancing areas is luminescence-based sensing and imaging[19–22]. By improving a detailed understanding of luminescence mechanisms, it is possible to design luminescent materials whose properties are susceptible to changes in the physical and chemical parameters of their environment, e.g. pressure[23–27] or temperature[20,28–32]. The increasing prevalence of scientific studies on luminescent sensors can be attributed to several factors, foremost among them the ability to remotely monitor the parameters being analyzed, i.e. band spectral position, intensity ratio, luminescence decay time or bandwidth. The optical excitation of phosphor and subsequent analysis of their spectroscopic properties eliminate the need for physical contact or the use of wires for electrical connection, offering unparalleled flexibility and versatility, opening new application possibilities not achievable with existing conventional contact techniques[33–35]. Notable among them is a real-time, contactless *in vivo* sensing of spectroscopic parameters[20,36,37]An additional key advantage of luminescent sensors lies in their adaptability: the performance of the material can be fine-tuned to meet the strict requirements of specific applications by modifying the host material composition [37–42]. This customization of novel, advanced materials enables not only localized spot sensing but also two-dimensional, real-time imaging. Importantly, luminescent materials allow for highly precise, accurate and rapid imaging of changes in the parameter of interest during system operation, without disrupting its functionality, which is an economically advantageous feature that simplifies workflows and reduces downtime[43–48].

Advanced bifunctional pressure and temperature sensors hold great potential across various fields, including biomedical sciences, aerospace, geothermal energy, and disaster prevention[49–53]. Their ability to monitor both parameters with high precision, especially when integrated with smart infrastructure and the Internet of Things, enhances efficiency and safety[54]. In biomedical sciences, these sensors can revolutionize patient monitoring by being integrated into wearable and implantable devices to track blood pressure, body temperature, or intracranial conditions, aiding early diagnosis and treatment. In aerospace, sensors embedded in aircraft wings or engines to simultaneously monitor pressure and temperature ensure structural integrity and thermal regulation in aircraft and spacecraft, preventing failures caused by extreme pressure or temperature fluctuations. In space exploration, such sensors can withstand harsh environments, providing vital data on external conditions and internal systems of spacecraft, ensuring mission success and astronaut safety. In geothermal energy, these sensors can optimize energy extraction and prevent equipment failure by monitoring subsurface conditions. In disaster prevention, optical bifunctional sensors can detect early signs of volcanic eruptions, earthquakes, and landslides, improving predictive modeling and risk mitigation. Their bifunctionality reduces costs, size, and complexity, making them practical for sustainable energy and disaster resilience. Additionally, they enhance control of industrial processes, environmental monitoring, and smart infrastructure by improving accuracy and reducing the need for multiple sensors. Their compact, multifunctional design makes them a versatile tool for advancing technology across multiple sectors.

Among the many uses of luminescent materials, luminescence thermometry stands out as a leading application[31,55–57]. This is evidenced by the generation of thousands of scientific publications on that topic in recent years, including numerous review papers[27,29,41,57–60]. The growing interest in luminescence thermometry is closely tied to the critical importance of temperature as a fundamental thermodynamic parameter in everyday life. The literature is full of reports on novel thermometric materials[29,61], innovative sensing strategies[62–66], and advancements designed to meet the demanding criteria of specific applications.

While luminescence thermometry dominates the field of optical sensing, the second-most critical thermodynamic parameter, i.e. pressure, can also be measured using luminescence [23,24,67–72]. Ruby ($Al_2O_3:Cr^{3+}$) remains the most widely used luminescent manometer, having served this purpose for decades[23,24,71,73,74]. However, recent years have shown significant advancements with new phosphors demonstrating promising sensing performance. Although luminescent materials for temperature and pressure sensing have traditionally been developed independently, it is important to recognize that spectroscopic properties of a single phosphor may exhibit susceptibility to changes of both parameters. This dual sensitivity can complicate measurements, as changes in one parameter can alter the calibration curve of the other (i.e. the well-known pressure-temperature cross-dependences)[75–77], leading to inaccuracies or biased readouts. Furthermore, in real-world applications where simultaneous fluctuations in pressure and temperature occur, using separate phosphors for each parameter may be impractical, especially in luminescence imaging scenarios that require the phosphor to coat the entire surface of the object being analyzed.

In response to these challenges, single- or multi-parameter bifunctional luminescent sensors have gained growing attention. These sensors offer not only significant scientific intrigue, but also substantial application potential. This review aims to critically examine bifunctional sensors reported in the literature, categorizing them into two main groups: (i) sensors that use the same spectroscopic parameter to detect both temperature and pressure and (ii) sensors that rely on two distinct luminescence parameters for simultaneous measurements of both state functions. In this review we discuss the advantages and limitations of each approach, providing a framework for selecting the most suitable solution based on specific application needs. Our analysis includes the latest advancements and material innovations in the field, offering a comprehensive overview of the state-of-the-art in multi-modal, bifunctional luminescence sensing.

## 2. Critical parameters for sensor performance determination

### 2.1 Absolute and relative sensitivity

To determine the applicability of potential bifunctional, optical pressure and temperature sensor it is crucial to indicate the performance of the sensing performance. One of the most important parameter that enables qualitative assessment the temperature/pressure induced changes of a given spectroscopic feature of the phosphor is the sensitivity of the sensor to given physical stimuli. In the literature two types of sensitivities are usually considered *i.e.* absolute sensitivity ($S_A$) and relative sensitivity ($S_R$). The $S_A$ parameter shows the absolute change of the measured optical parameter Δ (*e.g.* band shift or FWHM) per unit of physical parameter of pressure or temperature and is expressed as follows:

$$S_A = \frac{\partial \Delta}{\partial p, T} \qquad (1)$$

where $\Delta$ is a measured spectroscopic parameter (shift, LIR, lifetime, FWHM, etc.), and $p$ or $T$ are pressure or temperature, respectively. This parameter is very often used to describe the pressure or temperature sensors based on the spectral shift of the emission bands and less commonly used in the case of other approaches. To compare the performance of the sensor independently of the nature of material and measured parameter selected and optical setup features/configuration, the $S_R$ parameter expressing the relative change of the measured parameter per pressure or temperature unit (typically GPa or kbar for pressure and K for temperature) can be calculated as follows:

$$S_R = \frac{1}{\Delta}\frac{\partial \Delta}{\partial p, T} \cdot 100\% \qquad (2)$$

## 2.2 Temperature/pressure determination uncertainty

A third important parameter to the comparison of optical sensors is thermal/pressure resolution (or uncertainty), denoted as $\delta T$ or $\delta p$, respectively, which indicates the smallest value of physical parameters possible to detect using a particular sensor and experimental setup. Since this parameter depends on the uncertainty in the Δ determination (δΔ) the calculation of resolution can be done in two ways: I) empirically, by a large number of experimental measurements and calculations of standard deviation from the histogram, or II) by calculating it based on the signal-to-noise ratio and determined relative sensitivity, using Equation 3:

$$\delta p, T = \frac{1}{S_R}\frac{\delta\Delta}{\Delta} \tag{3}$$

## 2.3 Operating range

Another important parameter that describes the thermometric or manometric performance of the luminescence sensors is its operating range. This parameter determines the range of temperature or pressure in a given sensor can be used to reliably read a given physical quantity. Usually, this range refers to the range of variation of a given physical parameter in which monotonic changes in the spectroscopic parameter are observed. It is worth noting here that usually the width of the operating range is inversely proportional to the value of sensitivity because the high dynamic range of variation of a given physical parameter usually limits the operating range and, conversely, obtaining a wide operating range usually involves the use of materials with low sensitivity. This correlation is related to physical limitations of the variation of emission band intensity or luminescence kinetics. Some deviations from this correlation can be observed for spectral shift-based sensors.

## 2.4 Reproducibility

Other important factors of optical manometers and thermometers are repeatability and reproducibility. The repeatability can be examined by cycling measurement, *i.e.* repeatedly measured parameters in the initial and final boundary conditions, which reflects the stability of the material properties under given conditions. Reproducibility strictly depends on the equipment used for measurements and can be very individual for each laboratory. That is why the exact determination of the conditions is important in the development of novel bi-functional sensors.

### 2.5 Thermal invariability manometric factor and pressure invariability thermometric factor

In the characterization of conventional thermometers and luminescent manometers, the reciprocal effects of simultaneous variations in temperature and pressure are often overlooked. However, for the development of bifunctional luminescent sensors, it is essential to assess how changes in one parameter influence the sensitivity to the other. To address this, we propose a Thermal Invariability Manometric Factor (TIMF), defined as follows[78]:

$$TIMF = \frac{S_{R,p}}{S_{R,T}} \quad (4)$$

This parameter determines by how much the temperature of the system must change in order to observe changes in the spectroscopic parameter corresponding to a pressure change by unit

The analogous parameter is the Pressure Invariability Thermometric Factor (PITF):

$$PITF = \frac{S_{R,T}}{S_{R,p}} \quad (5)$$

which determines by how much the ambient pressure must change in order to observe changes in the spectroscopic parameter corresponding to a unit change in temperature

## 3. Mechanisms responsible for changes in spectroscopic parameters utilized for luminescence sensing

Although textbooks and review articles on luminescence thermometry and manometry often classify sensors into six categories - based on single-band intensity, bandwidth, polarization, spectral band shift, luminescence intensity ratio (LIR) of two emission bands, and luminescence kinetics - only the last three have significant practical value[28,79]. The limitations of other approaches stem from inherent dependencies on external factors or insufficient sensitivity[34,80]. For example, while single-band intensity can be sensitive to specific physical stimuli *e.g.,* $O_2$ concentration in pressure-sensitive paints (PSP)[81,82], it is also influenced by numerous external factors, such as the geometry of the detection system, the distance between the phosphor and the detector, optical excitation intensity, phosphor concentration, and many others[28,31,79]. These dependencies significantly undermine its reliability for practical applications. Similarly, the inherent sensitivity of bandwidth to temperature or pressure changes is typically low, necessitating the use of high-resolution spectral detectors, which complicates the measurement process. On the other hand, polarization-based luminescence is relatively rare and requires the use of polarizers, along with their adjustment during measurements. Additionally, the changes in polarization spectra are often minimal, limiting the precision and practicality of this approach.

Given these limitations, the discussion in this work will focus exclusively on the three spectroscopic parameters with substantial application value: spectral band shifts, LIR and luminescence kinetics. A critical aspect of designing sensor materials is a comprehensive understanding of the processes responsible for the observed changes in the spectroscopic properties of the phosphor. To this end, the following sections provide a focused overview of the most important mechanisms used for temperature and pressure sensing.

### 3.1 Bandshift

The spectral position of the luminescence band can shift due to various factors, including thermal and pressure effects[23,24,27,83]. Since the first report on a luminescent manometer in the 1970s, this parameter has been one of the dominant methods for pressure measurements, particularly in diamond anvil cells (DACs), where ruby ($Al_2O_3:Cr^{3+}$) remains the gold standard[23,24,71,73,84,85]. In contrast, this parameter is less frequently employed in luminescence thermometry, primarily due to its relatively low sensitivity to temperature changes compared to alternative approaches[28]. Despite this limitation, phosphors with spectrally narrow emission lines, such as lanthanide ions, can achieve exceptionally high precision in temperature readings in this approach. However, this method poses significant challenges for luminescence imaging, as it requires point-by-point spectral measurements across the emission band, a process that is time-consuming and demands high-resolution spectral detectors. Nonetheless, in isothermal or isobaric conditions, this approach remains viable and is actively utilized.

Electron-phonon coupling and interactions with host material phonons (both acoustic and optical branches) can significantly affect the spectral position of the emission band[86–89]. This is an effect, well described in the literature and observed for many phosphor materials. This effect affects the spectral position of the *R* band of $Cr^{3+}$ ions[24,71]. An increase in temperature can also lead to a change in the local environment of the luminescent ion through thermal expansion of the host material. Such an induced increase in bond length can cause a change in the spectral position of the band leading to its red-shift spectra. However, such induced changes will not always be isotropic. The anisotropy of the thermal expansion of the host material can lead to a thermally induced distortion of the local point symmetry of the luminescent ion leading to a change in the spectral position of the band.

For materials doped with transition metal ions, the energy levels are influenced by the crystal field strength (CFS), which depends on the metal-oxygen bond length and can be described using Tanabe-Sugano diagrams[90,91]. One of the most commonly used and thoroughly investigated representative of the transition metal ions used in sensing applications is $Cr^{3+}$ [92–94]. Therefore, the discussion presented below will be based predominantly on this ion as representative example. The CFS changes with the bond length ($R$) for $Cr^{3+}$ ions in octahedral symmetry proportionally to $R^{-5}$ [94]. Hence even minor changes in *R*, induced by the applied pressure, can significantly alter the CFS and, consequently, the energy levels of the ions. For spin-allowed transitions, such changes modify the energy separation between the ground and excited states, resulting in a shift in the spectral position of the emission band[16,92,94,95]. According to the corresponding Tanabe-Sugano diagram, compression of the material characterized by the weak or intermediate CFS typically causes a blueshift of the $^4T_2 \rightarrow ^4A_2$ emission band. The rate of this shift can reach up to 23 nm $GPa^{-1}$[96]. However, the broad nature of the emission band can limit the precision of measurements. A notable limitation of this approach arises when the crystal field splitting parameter (*Dq/B*) exceeds ~2.1 under compression[95,97]. Beyond this point, the energy of the $^4T_2$ level surpasses that of the $^2E$ level, causing the luminescence from $^4T_2$ state to vanish. Conversely, at very low *Dq/B* values under ambient pressure, nonradiative depopulation processes of the $^4T_2$ level can reduce luminescence intensity.

On the other hand, pressure evolution of the spin-forbidden electronic transitions, such as $^2E \rightarrow ^4A_2$, is governed by an alternative mechanism. In this case, the energy of the $^2E$ level is independent of the CFS, but compression still affects luminescence through enhanced metal-oxygen bond covalency (nephelauxetic effect), resulting in a slight redshift of the emission band[98–100]. For ruby-based manometers, this shift has been quantified as 0.365 nm $GPa^{-1}$ and is reliable and monotonic over a wide range of pressures, from ambient up to 150 GPa[24,74]. Similar phenomena are observed in systems involving *d*-*f* or charge transfer transitions. In these cases, the spectral shifts are driven by changes in the configuration of the excited state or inter-ion distances between interacting ions. These mechanisms further expand the utility of spectral position shifts for both temperature and pressure sensing in luminescent systems. Additionally, the applying pressure to luminescent materials can influence both the energy levels of luminescent centers and the band gap between the valence and conduction bands[101–103]. This pressure-induced modulation can lift the degeneracy of energy levels, leading to alterations in emission spectra. In specific instances, such as $Y_3Al_2Ga_3O_{12}:Ce^{3+}$, this effect results in the emergence of $Ce^{3+}$ luminescence, followed by a gradual spectral shift due to changes in the relative intensities of *d*-*f* transitions in the observed emission spectra[103].

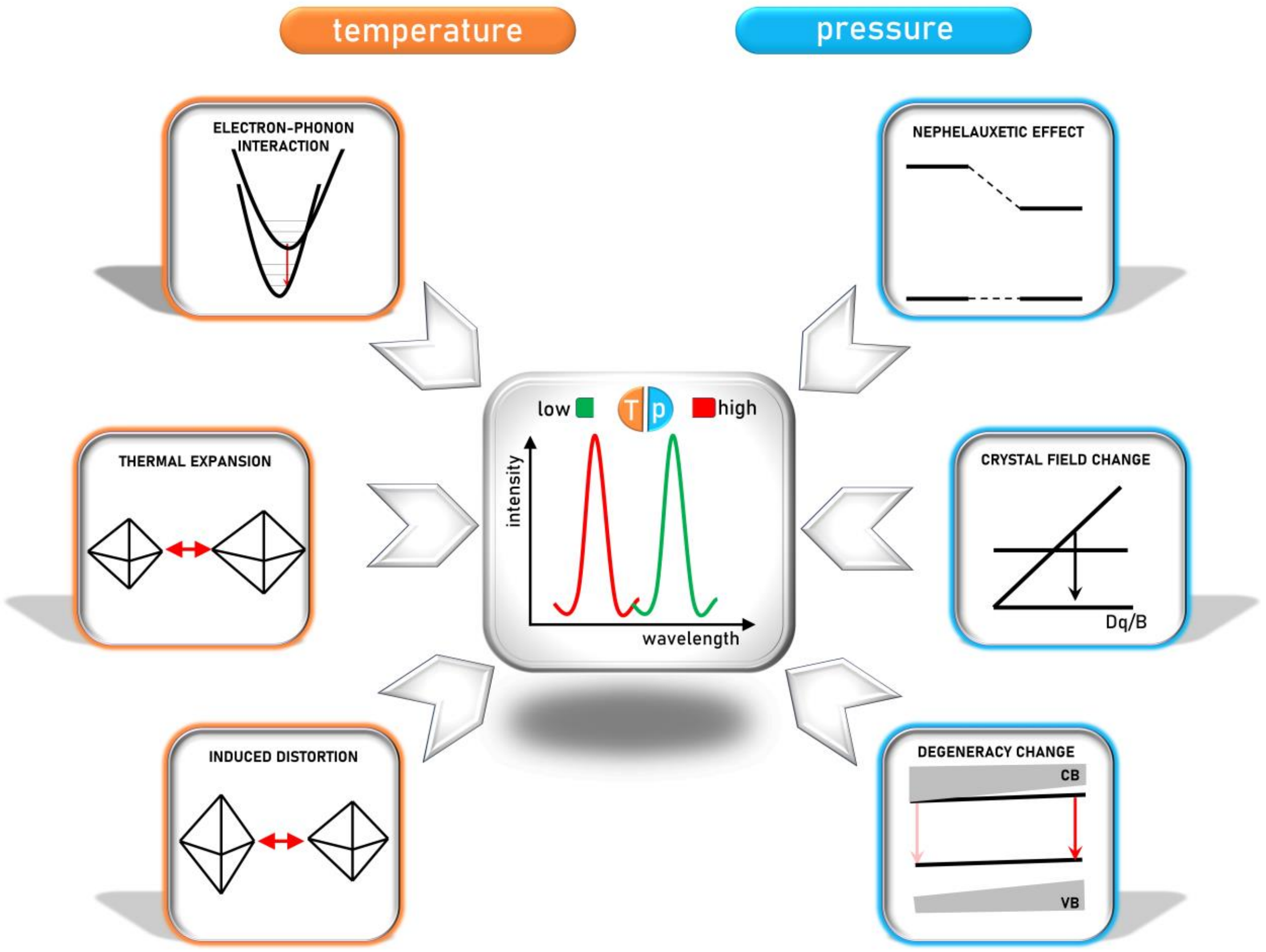

**Figure 1.** Different mechanisms affecting the emission band shift under change of temperature and pressure.

## 3.2 Luminescence Intensity Ratio

The luminescence intensity ratio (LIR) is among the most widely utilized parameters for temperature and pressure sensing[27,29,41]. Unlike methods based on the absolute intensity of a single emission band, the ratiometric approach compares the changes in one emission band against a reference band. This methodology inherently reduces the influence of external factors such as system geometry, material concentration, and optical excitation power density, enabling robust and accurate measurements. Moreover, it provides exceptionally high sensitivity to variations in the physical quantity being measured. Notably, this approach can be easily adapted for luminescence imaging by the use of spatial detectors with appropriate optical filters, or by assigning the analysed emission bands to different RGB channels of a sensor[104–106]. These advantages have resulted in a proliferation of luminescence-based thermometers in the literature, as well as boosted application of ratiometric techniques in luminescence manometry[31,55,60,69,83,107–109].

Several key mechanisms influence thermally induced changes in the LIR. One of the most common is thermalization. When two or more energetically proximate excited emitting levels (typically separated by less than 2000 $cm^{-1}$) exist, an increase in temperature redistributes the population between these levels[57]. This redistribution, described by the Boltzmann distribution, leads to predictable changes in LIR. Such luminescence thermometers, termed primary thermometers, offer universal calibration curves governed by fundamental Boltzmann statistics ensuring consistency across various operating conditions, according to the Boltzmann equation[110,111]. This single-ion approach involves emission levels originating from the same luminescent center.

When employing multiple luminescent dopants or emission centers, the changes of the LIR values can result from differences in the thermal depopulation rates of the excited states associated with these centers, as well as variations in the probabilities of energy migration to traps and defects states[112–119]. In such cases, dopants with distinctly different energy levels or electronic configurations are typically used[31,59,60,120]. For example, ions with varying probabilities of nonradiative multiphonon depopulation, governed by energy gaps rule, to lower energy levels, are often utilized[121]. Additionally, combinations of lanthanides and transition metals in a single sensor material are commonly found, as their differing electronic transition mechanisms and sensitivities to environmental changes can enhance the performance of the resulting luminescent sensors. These approaches yield high relative sensitivities, often surpassing those achieved with thermalization[122–127]. However, they do not constitute primary thermometers, sometimes complicating the theoretical description of internal mechanisms and calibration curves.

Another mechanism that influences thermal changes of the LIR parameter is nonresonant temperature-induced energy transfer, which involves phonon-assisted processes between energy levels or luminescent ions[128–133]. This mechanism often provides higher sensitivities than thermalization while retaining some degree of theoretical predictability, as described by the Miyakawa-Dexter model[134].

In the case of material compression, entirely different mechanisms influence the LIR parameter. For materials doped with $Ln^{3+}$ ions, pressure primarily activates additional energy transfer processes, due to the shortened interionic distances in the compressed structures[135–142]. For example, enhanced cross-relaxation processes can result in nonradiative depopulation of one of the emitting levels, as observed in $Er^{3+}$-doped materials[141]. In the case of transition metal ions, pressure-induced modifications to the crystal field strength alter the energy of electronic levels[135,143–151]. This change can affect the population distributions *via* mechanisms such as thermalization. For instance, in $Cr^{3+}$-doped materials, compression results in enhanced energy of the $^4T_2$ level, altering the luminescence intensity ratio of the transitions $^2E \rightarrow {}^4A_2$ and $^4T_2 \rightarrow {}^4A_2$[90,94].

Additionally, the pressure-dependence of transition metal ion energy levels can be exploited in ratiometric manometry by using co-doped $Ln^{3+}$ ions as a luminescent reference. Similar effects are observed in emission bands associated with *d-f*, charge transfer, or singlet-triplet transitions, albeit through different mechanisms. Such changes often result in spectral shifts, as described earlier, but they can also serve as a basis for ratiometric pressure measurements. For instance, by analyzing two spectral regions of a single emission band rather than distinct bands, a highly sensitive luminescence manometer can be developed. This approach offers some advantages, which will be discussed in detail later in this paper. Compression of the material may modify the energy gap between valence and conduction band[152–154]. Since the energy separation between some emitting states and the bottom of the conduction band may affect the probability of the depopulation of the emitting state and thus its emission intensity, this effect can be exploited in the luminescence manometry as well. When the additional emission band not affected by this process will be used as a luminescence reference the ratiometric luminescent manometer can be developed[139].

In summary, a variety of physical mechanisms-ranging from thermalization of excited states and energy transfer processes to pressure-induced modifications affects LIR changes, enabling their use in remote sensing and imaging of temperature and pressure. These mechanisms provide a versatile framework for developing advanced luminescence-based sensors and imaging systems for diverse applications.

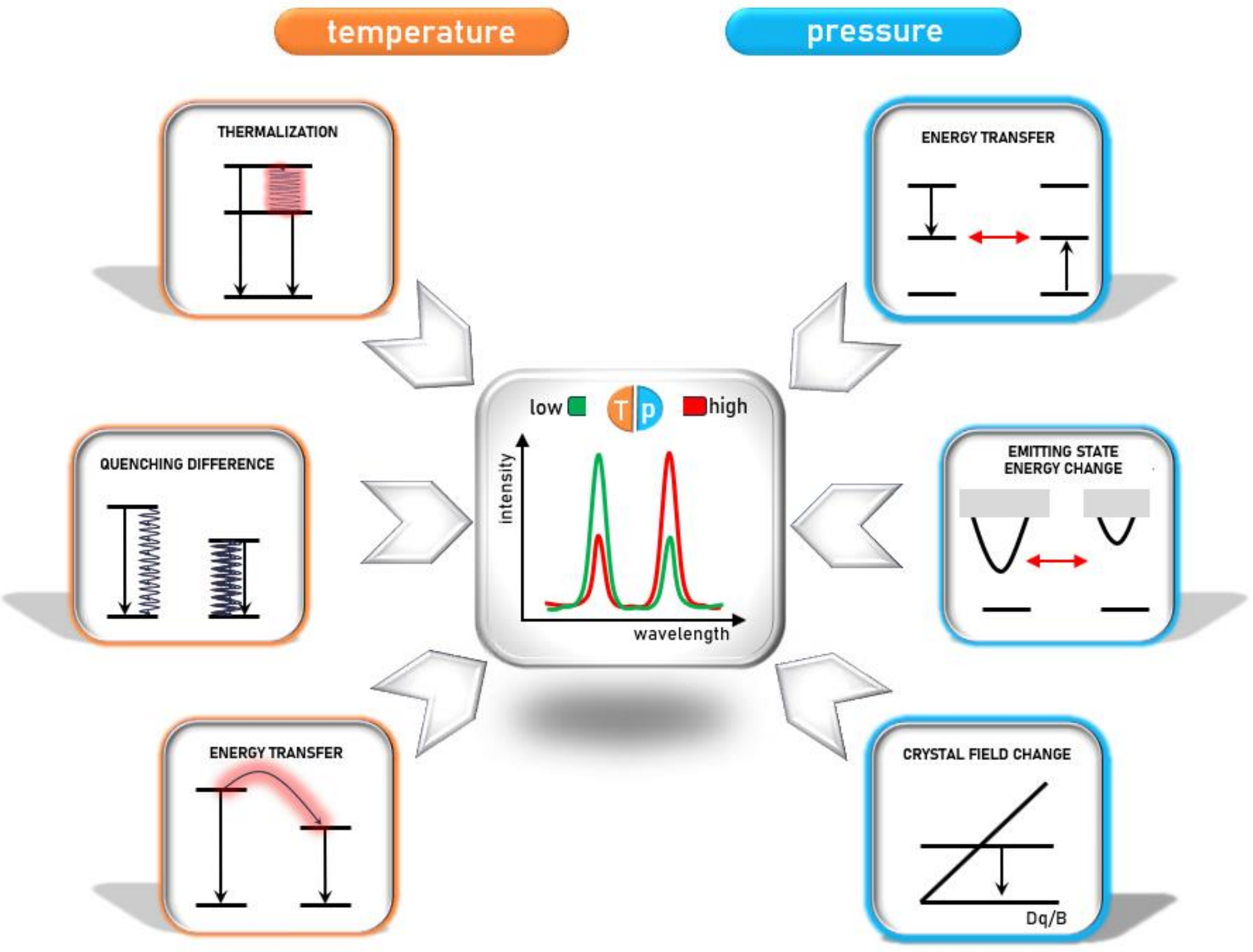

**Figure 2.** Different mechanisms affecting the luminescence intensity ratio under change of temperature and pressure.

### 3.3 Luminescence kinetics

The luminescence kinetics of dopant ions exhibit high sensitivity to temperature variations, primarily due to the thermal depopulation of emitting levels. In lanthanide ions, the predominant nonradiative depopulation mechanism of excited states involves multiphonon processes[155–158]. According to the energy gap rule, a smaller energy separation between the emitting level and the next lower level increases the probability of nonradiative transitions[121]. The strong temperature dependence of the probability of this process enables its application in temperature sensing[134]. Furthermore, by engineering the host material, the maximum phonon energy can be modulated, thereby influencing the operational range and sensitivity of lifetime-based luminescence thermometers[128,156,159].

Transition metal ions offer additional flexibility in modulating the probability of nonradiative depopulation processes. This is attributed to the involvement of intersection points between the parabolas of excited and emitting states in the depopulation process[94,95,99,160]. Since the energy levels of transition metal ions can be adjusted by altering the crystal field strength, the activation energy (energy difference between the bottom of the emitting state parabola and the intersection point) can also be modulated[42,161]. Electron-phonon interactions further allow tuning of the activation energy required for non-radiative depopulation of the emitting level[162–164]

Phonon-assisted nonradiative depopulation is another prevalent mechanism utilized in lifetime-based luminescence thermometry[165–168]. When a small energy mismatch between excited levels of two dopant ions occurs, energy transfer can occur with phonon absorption or emission, with the probability of this process increasing with temperature as the number of phonons needed to bridge the energy gap decreases[167]. Careful selection of host materials (considering phonon energy) and interacting co-dopant ions allows for regulation of the thermometric performance of such temperature sensors[130,169,170].

Compression of phosphors leads to a reduction in the distance between dopant ions, facilitating energy transfer between them. In this context, cross-relaxation is a significant process that notably affects the luminescence kinetics of lanthanide ions[109,139,141,171]. Defined as a resonant process, its probability is typically independent of temperature changes (however, some reports on thermally activated cross-relaxation can be found in the literature). However, reducing the distance between dopant ions enhances cross-relaxation, thereby shortening the luminescence decay profile - a phenomenon that can be exploited in luminescence manometry[139–141]. It is important to note that the shortening of lifetimes of excited states in lanthanide doped phosphor is a common effect of temperature variations, which may complicate the identification of the dominant physical parameter influencing changes in luminescence kinetics.

An intriguing effect from an application standpoint is the alteration of luminescence kinetics in transition metal ions due to changes in crystal field strength associated with material compression. The spin-orbit interaction between the excited states of transition metal ions and the overlap of their wavefunctions make luminescence kinetics highly sensitive to pressure changes[92,95,99,160]. Notably, in ions such as $Cr^{3+}$, where spin-allowed and spin-forbidden states interact (as in $3d^3$ ions), the small energy difference between the $^2E$ ($^2E \rightarrow {}^4A_2$ spin-forbidden transition) and $^4T_2$ ($^4T_2 \rightarrow {}^4A_2$ spin-allowed transition) levels, coupled with the strong dependence of the $^4T_2$ level energy on crystal field strength, results in changes in luminescence kinetics[172–182]. In phosphors with intermediate and strong crystal fields, material compression increases the energy of the $^4T_2$ level, decreasing the coupling strength between them, leading to an increase in the lifetime of the $^2E$ level[172–175,177,181]. A similar effect is observed in weak crystal field phosphors, where emission from the $^4T_2$ level exhibits prolonged luminescence kinetics due to increased overlap of the $^2E$ and $^4T_2$ wave functions under applied pressure[174]. This pressure-induced prolongation of luminescence kinetics contrasts with the typical effects observed with increasing temperature and holds significant potential for sensory applications.

Additionally, the Jahn-Teller effect can influence the luminescence kinetics of dopant ions, with its consequences being particularly pronounced in transition metal ions[183–186]. In host materials exhibiting anisotropic compression, applied pressure can distort the crystallographic site occupied by the dopant ion, affecting bond covalency and modifying the selection rules for electronic transitions. Consequently, changes in luminescence kinetics are observed; however, these changes are generally smaller compared to those resulting from alterations in crystal field strength.

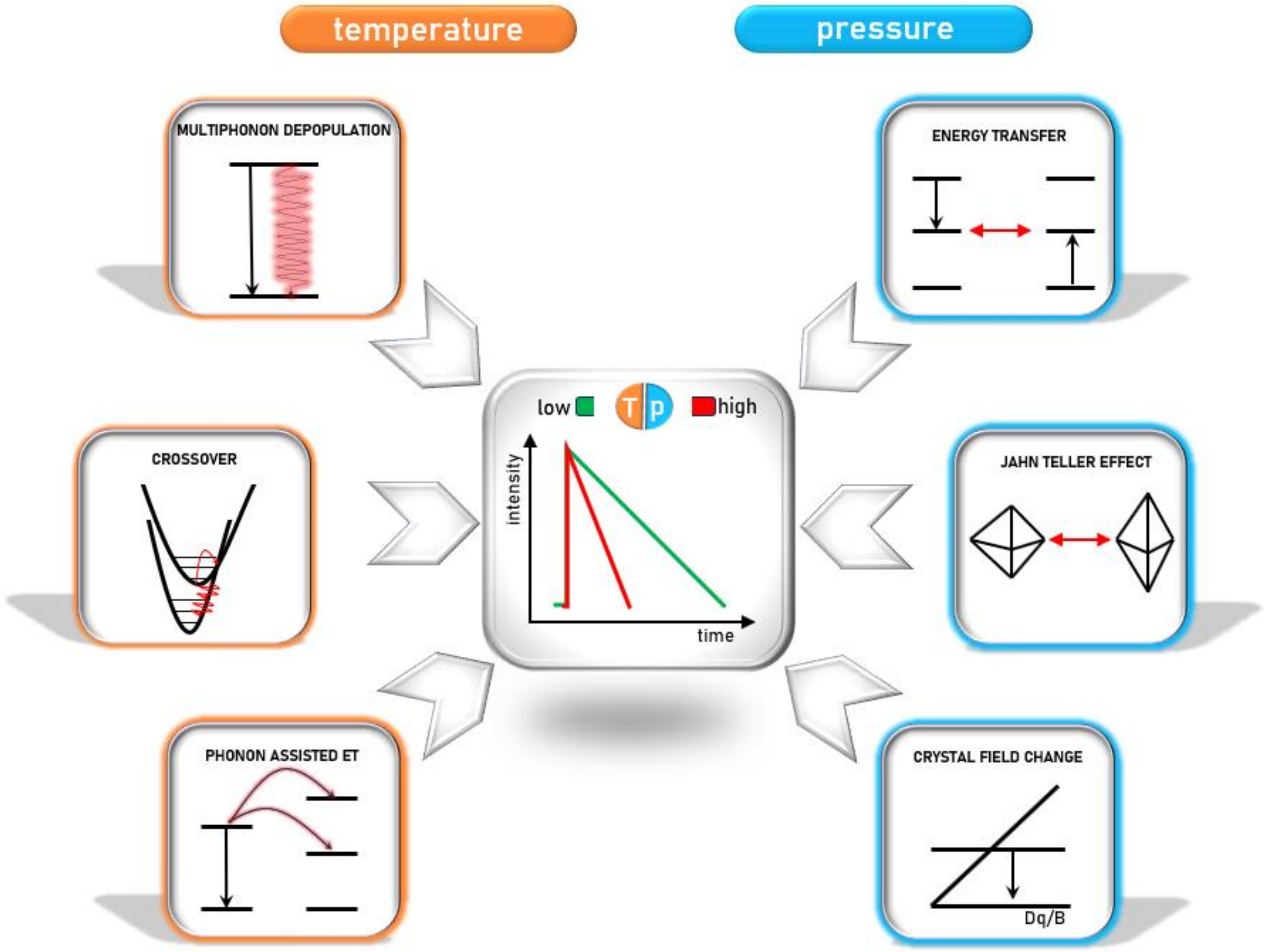

**Figure 3**. Different mechanisms affecting the luminescence kinetics under change of temperature and pressure.

Finally, concerning the use of bandwidth as a thermometer or manometric parameter, one should be conscious of its severe limitations and inherent issues. Specifically, determining the FWHM in the case of a spectrum that is neither a single Lorentzian nor a Gaussian is highly questionable and unjustified from a physical perspective. Of course, in such a case, deconvolution can be performed, but it is always highly uncertain because it results in multiple interdependent parameters. However, there are some reports available in the literature dealing with FWHM as a sensing parameter, even in the context of bi-functional temperature-pressure sensors. Hence, we briefly describe this approach here. In the case of pressure, the evolution of the bandwidth is associated mainly with the I) pressure-induced increase of the crystal-field strength, leading to the larger band splitting; II) enhanced electron−phonon coupling; III) increasing strains/distortions of the materials, formation of crystal defects; and IV) deviations from hydrostaticity observed at higher pressure values. On the other hand, the thermal broadening of the band is primarily governed by I) the static contribution $\Delta E_{st}(T)$ caused by the changes in the geometry of the site occupied by the optically active ion in the crystal, triggered by the lattice thermal expansion; and II) the vibrational contribution $\Delta E_{vib}(T)$ related to the electron−phonon interactions.

**4. Bifunctional sensors**

The bifunctional sensors of pressure and temperature can be classified into two main categories, i.e. 1) as sensors allowing detection of both quantities operating with the same spectroscopic parameter (single-parameter bifunctional sensors), or 2) using a distinct temperature-independent parameter for pressure determination and *vice versa*. As expected, the first category is much more common, and frequently reported in the literature, mainly due to the simplicity of the operating principle and mutual dependence of the vast majority of luminescence features alike on pressure and temperature[69,71,175,187–192]. However, such sensors do not provide reliable pressure detection at variable temperature conditions and *vice versa*. Whereas, in the case of the second category, there are much less report on such truly bifunctional sensors, however, they are much more reliable for sensing at extreme conditions of both pressure and temperature[25,72,140,193–200].

**4.1 Single-parameter bifunctional sensors**

Probably, the first reported bifunctional sensors of pressure and temperature is also ruby. Independently in 1999 Rekhi et al.[188] as well as Gibson et al.[71] used the temperature dependence of the $R_1$ and $R_2$ emission lines spectral shifts of $Cr^{3+}$ in ruby single microcrystals and films ($S_A$ ≈0.007 nm $K^{-1}$) for temperature sensing applications[23,24,71]. The researchers confirmed the possibility of temperature sensing up to 873 K under high-pressure conditions of 15 GPa in a DAC chamber, using the same approach (line shift) as for pressure sensing[23,24].

In 2018 Romanenko et al.[189] reported the use of $SrB_4O_7:Sm^{2+}$ for sensing of both quantities. The authors calibrated spectral shifts of several narrow emission lines of $Sm^{2+}$ as a functions of pressure and temperature. All of the observed emissions lines red-shift with pressure at different rates, and in the case of temperature alike red- and blue-shifts were observed. The authors claimed that analysing the multiple parameters (several lines) it is possible to extract their pressure and temperature contributions using the appropriate cross-calibrations procedures and the proposed algorithms, in order to monitor both pressure and temperature under extreme conditions (up to 7 GPa and 723 K), with standard deviation of 8 K.

On the other hand, utilizing alike $Eu^{2+}$ and $Sm^{2+}$ emission in the same strontium borate host material, enables to perform ratiometric pressure sensing, with maximal sensitivity of 13.8 % $GPa^{-1}$[190] . In that case, the integrated intensities of the $^6P_{7/2} \rightarrow {}^8S_{7/2}$ (UV emission) and $^5D_0 \rightarrow {}^7F_0$ (visible emission) transition of $Eu^{2+}$ and $Sm^{2+}$, respectively, were used and their intensity ratio was correlated with pressure using a polynomial fit. The operating range of that sensor is between ≈10 and 40 GPa. Moreover, the developed pressure gauge also allows optical detection of the local temperature using two different approaches, namely a ratiometric and lifetime-based modes. In the case of ratiometric sensing, temperature can be detected using either LIR of the UV emission of $Eu^{2+}$ (alike 4f-4f and 5d-4d transitions) in respect to the $Sm^{2+}$ emission in the visible range.

Another report about the bifunctional sensors based on the same borate host is about $Tm^{2+}$-doped $SrB_4O_7$[69]. This material has a broad emission band centred around 600 nm, associate with allowed, inter-configurational transition of $Tm^{2+}$ ($4f^{12}5d^1 \rightarrow 4f^{13}$). In this report the band shift and its FWHM were used as manometric parameters (calibrated up to ≈13 GPa), both allowing alike pressure and temperature determinations, resulting in maximal pressure sensitivities of $S_A (\lambda)$ ≈0.43 nm $GPa^{-1}$ and $S_A (FWHM)$ ≈0.89 nm $GPa^{-1}$, respectively.

The comparison of the sensing performance of different materials reported as bifunctional temperature and pressure sensors operating with a single-parameter is presented in Table 1.

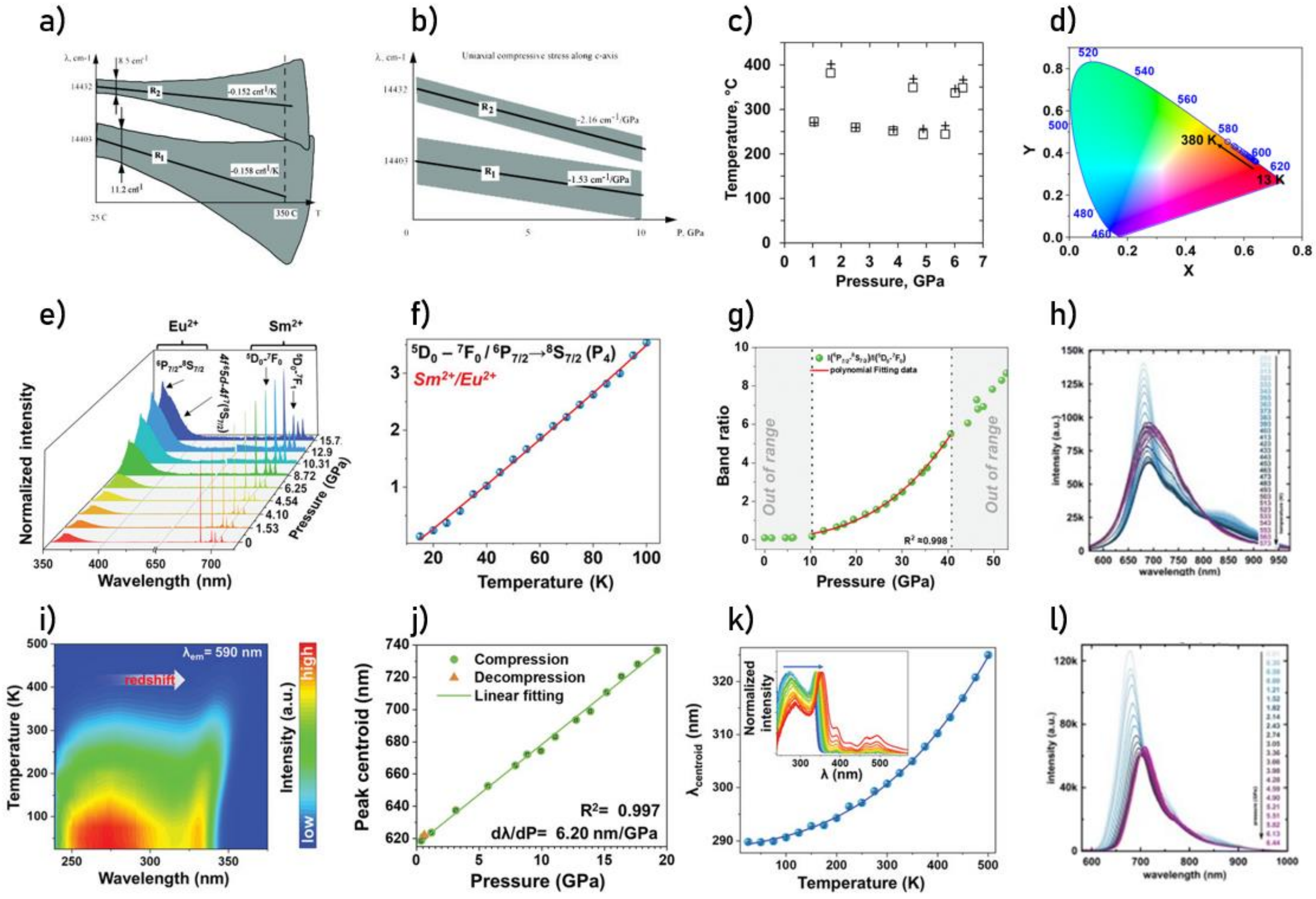

**Figure 4.** Thermal - a); and pressure dependence of the energies of the emission lines of $Cr^{3+}$ in $Al_2O_3$:$Cr^{3+}$- b)[71]; Pressure and temperature calculated using cross-calibration of $\lambda_1$ of $SrB_4O_7$:$Sm^{2+}$and ruby fluorescence lines (squares), and $\lambda_1 + \lambda_2$ $SrB_4O_7$:$Sm^{2+}$ fluorescence lines (crosses)- c)[189]; thermal dependence of the chromatic coordinates for $SrB_4O_7$:$Tm^{2+}$- d)[69]; pressure-dependent emission spectra of the $SrB_4O_7$:0.01$Eu^{2+}$, 0.03$Sm^{2+}$ phosphor- e)[190]; thermal - f) and pressure dependence of LIR for $SrB_4O_7$:0.01$Eu^{2+}$, 0.03$Sm^{2+}$ phosphor - g)[190]; the influence of temperature on the emission spectra of PTCDI-Ph - h)[192]; thermal dependence of excitation spectra of ZnS/CaZnOS:$Mn^{2+}$ - i)[191]; the influence of the pressure on the $Mn^{2+}$ band position - j)[191]; the influence of the temperature on the band position in the excitation spectra of ZnS/CaZnOS:$Mn^{2+}$ - k)[191]; pressure dependence of the emission spectra of PTCDI-Ph - l)[192].

**Table 1.** The list of the bifunctional luminescent sensors operating with a single spectroscopic parameter.

| Material | Thermometric parameter | $S_{R,T}$ or $S_{A,T}$ | Manometric parameter | $S_{R,p}$ or $S_{A,p}$ | Reference |
|---|---|---|---|---|---|
| Ruby–$Al_2O_3:Cr^{3+}$ | Bandshift: $R_1$ line ($^2E \rightarrow ^4A_2$) $Cr^{3+}$ | 0.007 nm $K^{-1}$ | Bandshift: $R_1$ line $^2E \rightarrow ^4A_2$ $Cr^{3+}$ | 0.365 nm $GPa^{-1}$ | [23,24,188] |
| $SrB_4O_7:Sm^{2+}$ | Bandshift:<br>$^5D_0 \rightarrow ^7F_0$ (0-0) $Sm^{2+}$<br>$^5D_0 \rightarrow ^7F_1$ (0-1) $Sm^{2+}$<br><br><br>$^5D_0 \rightarrow ^7F_2$ (0-2) $Sm^{2+}$ | <br>2.00 nm $K^{-1}$<br>0.0907 nm $K^{-1}$<br>0.285 nm $K^{-1}$<br>1.82 nm $K^{-1}$<br>0.192 nm $K^{-1}$<br>0.0682 nm $K^{-1}$<br>0.429 nm $K^{-1}$<br>0.110 nm $K^{-1}$ | Bandshift:<br>$^5D_0 \rightarrow ^7F_0$ (0-0) $Sm^{2+}$<br>$^5D_0 \rightarrow ^7F_1$ (0-1) $Sm^{2+}$<br><br><br>$^5D_0 \rightarrow ^7F_2$ (0-2) $Sm^{2+}$ | <br>0.238 nm $GPa^{-1}$<br>0.209 nm $GPa^{-1}$<br>0.292 nm $GPa^{-1}$<br>0.228 nm $GPa^{-1}$<br>0.307 nm $GPa^{-1}$ (0-10 GPa) | [189] |
| | | | | 0.228 nm $GPa^{-1}$ (10-60 GPa)<br>0.183 nm $GPa^{-1}$<br>0.255 nm $GPa^{-1}$<br>0.452 nm $GPa^{-1}$ | |
| $SrB_4O_7:Eu^{2+}-Sm^{2+}$ | ratiometric<br>and lifetime-based modes; both 4f-4f and 5d-4f<br>Lifetime: $^5D_0$ $Sm^{2+}$ | <br><br><br>3.17 % $K^{-1}$ | LIR: $^6P_{7/2} \rightarrow {}^8S_{7/2}$ $Eu^{2+}$/$^5D_0 \rightarrow {}^7F_0$ $Sm^{2+}$ | 13.8 % $GPa^{-1}$ | [190] |
| $SrB_4O_7:Tm^{2+}$ | Bandshift: $4f^{12}5d \rightarrow 4f^{13}$ $Tm^{2+}$<br>FWHM: $f^{12}5d \rightarrow 4f^{13}$ $Tm^{2+}$<br>Lifetime: $4f^{12}5d$ $Tm^{2+}$ | 1.44 $cm^{-1}$ $K^{-1}$<br>3.85 $cm^{-1}$ $K^{-1}$<br>4.16% $K^{-1}$ | Bandshift: $f^{12}5d \rightarrow 4f^{13}$ $Tm^{2+}$<br>FWHM: $f^{12}5d \rightarrow 4f^{13}$ $Tm^{2+}$ | 0.43 nm $GPa^{-1}$<br>0.89 nm $GPa^{-1}$ | [69] |
| $K_2Ge_4O_9:Mn^{4+}$ | Bandshift:R-line $^2E\rightarrow^4A_2$ $Mn^{4+}$ | 0.02 nm $K^{-1}$ | Bandshift: R-line $^2E\rightarrow^4A_2$ $Mn^{4+}$<br>LIR: $^2E\rightarrow^4A_2$ $Mn^{4+}$/$^2E\rightarrow^4A_2$ $Mn^{4+}$Lifetime:$^2E$ $Mn^{4+}$ | 0.59 nm $GPa^{-1}$<br>21.7 % $GPa^{-1}$<br>12 % $GPa^{-1}$ | [175] |
| ZnS/CaZnOS: $Mn^{2+}$ heterostructure | Bandshift:$^4T_1\rightarrow^6A_1$ $Mn^{2+}$<br>FWHM: $^4T_1\rightarrow^6A_1$ $Mn^{2+}$<br>Lifetime: $^4T_1$ $Mn^{2+}$ | 0.015 nm $K^{-1}$<br>0.057 nm $K^{-1}$<br>0.4% $K^{-1}$ | Bandshift: $^4T_1\rightarrow^6A_1$ $Mn^{2+}$ | 6.2 nm $GPa^{-1}$ | [191] |
| PTCDI-Ph (phenyl-substituted perylene diimides) | Bandshift: π-π stacking interactions | 0.068 nm $K^{-1}$ | Bandshift: π-π stacking interactions | 8.33 nm $GPa^{-1}$ | [192] |
| $Lu_2Mg_2Al_2Si_2O_{12}:Bi^{3+},Tb^{3+},Eu^{3+}$ | LIR:<br>$^5D_4\rightarrow^7F_5$ $Tb^{3+}$/$^5D_0\rightarrow^7F_2$ $Eu^{3+}$<br>$^5D_4\rightarrow^7F_5$ $Tb^{3+}$/$^5D_0\rightarrow^7F_4$ $Eu^{3+}$ | 0.15 % $K^{-1}$<br>0.21 % $K^{-1}$ | $^5D_4\rightarrow^7F_5$ $Tb^{3+}$/$^5D_0\rightarrow^7F_2$ $Eu^{3+}$ | 16% $GPa^{-1}$ | [187] |

## 4.2 Multi-parameter bifunctional sensors

To the best of our knowledge, the first bifunctional sensor of both discussed physical quantities, i.e. temperature and pressure, was reported in 2012 by León-Luis et al[201]., who showed the possibility of pressure and temperature sensing using luminescence of $Nd^{3+}$ ions embedded in gadolinium scandium gallium garnet crystals, i.e. $Gd_3Sc_2Ga_3O_{12}$:$Cr^{3+}$, $Nd^{3+}$. Authors used line shifts of the $R_{1,2}(^4F_{3/2})\rightarrow Z_5(^4I_{9/2})$ transitions for pressure sensing ($S_A$ = −11.3 $cm^{-1}$ $GPa^{-1}$) for $R_1$ and −8.8 $cm^{-1}$ $GPa^{-1}$ for $R_2$) up to 12 GPa. Whereas the LIR of the $R_{1,2}(^4F_{3/2})\rightarrow Z_1(^4I_{9/2})$ transitions, i.e. $R_1/R_2$ was used for temperature sensing in the cryogenic temperature range (10-300 K), with sensitivity reaching 6.7 % $K^{-1}$ around 20 K. Importantly, the observed $R_{1,2}\rightarrow Z_5$ emission lines of $Nd^{3+}$ barely shift with temperature, so monitoring their spectral shift guarantee reliable pressure determination. However, the relative intensities of the $R_{1,2}\rightarrow Z_1$ emission bands also change with pressure, hampering temperature determination under extreme condition of high pressure.

Another approach involves up-converting nanoparticles of $LaPO_4$ co-doped with $Yb^{3+}$ and $Tm^{3+}$ ions for ratiometric temperature sensing based on $Tm^{3+}$ thermally coupled levels $^3F_{2,3}/^3H_4\rightarrow{}^3H_6$ up to 773 K, and pressure sensing based on band shift of the $Tm^{3+}$ $^3H_4\rightarrow{}^3H_6$ transition ($S_A$ = 0.25 nm $GPa^{-1}$) or the band intensity ratio of $Tm^{3+}$ $^3H_4/^1G_4\rightarrow{}^3H_6$ transitions ($S_R$ <10% $GPa^{-1}$) up to 25 GPa[140]. Both methods of pressure sensing ensured rather rough estimation of the pressure values of around ±1 GPa. Due to the non-negligible impact of temperature on both manometric parameters, the authors were unable to determine pressure under elevated temperature conditions. However, authors successfully detected temperature changes under high-pressure conditions with thermal resolution of ≈4 K, i.e. in a DAC chamber, where the material was simultaneously compressed (up to ≈5 GPa) and heated up to ≈475 K.

For dual sensing, $NaBiF_4:Yb^{3+},Er^{3+}$ nanoparticles leverage well-resolved Stark components of $Er^{3+}$ NIR emission ($^4I_{13/2} \rightarrow {}^4I_{15/2}$) at ~1500 nm for pressure detection with a sensitivity of 0.8 nm $GPa^{-1}$. For pressure sensing, the authors used spectral shift of the well-resolved and narrow bands of $Er^{3+}$ NIR emission located around 1500 nm, associated with Stark sub-levels of the $^4I_{13/2} \rightarrow {}^4I_{15/2}$ transition of $Er^{3+}$, with sensitivity reaching up to $S_A \approx 0.8$ nm $GPa^{-1}$ for one of the crystal field component. It is worth noting, that the mentioned 4*f*-4*f* emission band, including the Stark components, barely shift with temperature, ensuring reliable pressure readouts under elevated temperature conditions. Whereas for temperature sensing the authors proposed commonly used procedure in luminescence thermometry, i.e. LIR of the two thermally coupled $^2H_{11/2}/^4S_{3/2}$ levels of $Er^{3+}$ ions. Unfortunately, the mentioned LIR parameter was also affected by pressure, changing from 0.29 at ambient pressure to 0.21 at ≈13 GPa, resulting in pressure sensitivity of ≈2%$GPa^{-1}$, and hampering reliable temperature detection under extreme condition of pressure. In both case the 980 NIR laser was used for excitation, resulting in down-shifting NIR emission (above ≈1000 nm) and up-conversion luminescence below ≈900 nm. Please note, that in this report the cross-calibration pressure-temperature experiments were not performed[200].

A complementary approach involves up-converting $YF_3:Yb^{3+},Er^{3+}$ micron-sized particles[199]. The authors also use thermally coupled $Er^{3+}$ levels, located in a green spectral region. However, in that case the material studied revealed a negligible and random change of the mentioned LIR parameter with pressure, at least up to 10 GPa, allowing remote temperature monitoring under high-pressure conditions. On the other hand, for pressure sensing, the authors used one of the Stark sub-levels of the $Er^{3+}$ ions $^4F_{9/2} \rightarrow {}^4I_{15/2}$ transition (red upconversion emission centered around 660 nm), with a sensitivity of ≈0.2 nm $GPa^{-1}$. Noteworthy, this narrow band has only a very low and well-defined temperature dependence of ≈0.004 nm $K^{-1}$, allowing reliable optical pressure detection under variable temperature condition. In that case, pressure sensing accuracy can be simply improved by applying the determined temperature-corrected calibration curve, which is especially important for measuring at extreme temperature conditions, where the system is significantly heated/cooled (i.e. by ~$10^1$-$10^2$ K). Importantly, the performed cross-calibration experiments of the simultaneously compressed and heated system (sensor material placed inside the DAC chamber located in the tubular furnace), confirmed the possibility and reliability of remote detection of alike pressure and temperature under extreme conditions of both factors. Using ruby crystals and thermocouple as reference sensors, the accuracies of pressure and temperature sensing under extreme conditions (up to 8 GPa and 475 K) were estimated to ≈0.2-0.3 GPa and ≤ 5 K, respectively.

In 2021 Sojka et al[72]. reported the use of $Pr^{3+}$-doped $Y_2Ge_{0.1}Si_{0.9}O_5$ material for band-shift luminescence manometry and LIR-based thermometry. In this case, pressure detection was realized analyzing spectral shift of the 5*d*→4*f* emission band of $Pr^{3+}$ ($S_A$ ≈1.28 nm $GPa^{-1}$) or its 4f→5d excitation band ($S_A$ ≈0.43 nm $GPa^{-1}$), centered around 320 nm and 250 nm, respectively. Whereas for temperature sensing, various LIRs between the 5*d*→4*f* and 4*f*→4*f* emission of $Pr^{3+}$ were used. However, the authors did not perform any cross-calibration experiments, i.e. they did not examine the system which is simultaneously compressed and heated. In fact, it seems that alike pressure affected the mentioned LIR parameters, and temperature affected the band centroids of the inter-configurations 5*d*-4*f* transitions used.

It is worth noting, that $SrB_4O_7:Sm^{2+}$ might be considered as a promising sensor of pressure and temperature working under extreme conditions, operating not only with a single parameter (i.e. emission line shift) but also with different parameters. For pressure sensing under variable thermal conditions, the 4*f*-4*f* emission line shift ($^5D_0 \rightarrow ^7F_0$) of $Sm^{2+}$ can be used, which is temperature-independent according to several reports[202,203] . Whereas, for temperature monitoring, one could use LIR of its 5*d*-4*f*/4*f*-4*f* emission, as reported by Cao et al[204]. However, it is still unclear how pressure would affects the relative intensities of the mentioned inter- and intra-configurational transitions, and if the reliable temperature sensing could be achieved under high-pressure conditions.

Phosphors doped with transition metal ions, particularly $Cr^{3+}$ and $Mn^{4+}$, are extensively studied due to their intense luminescence and high sensitivity of their optical properties to environmental changes[27,78,96,172,173,205]. For instance, in $K_2Ge_4O_9$:0.1%$Mn^{4+}$, a thermal shift of the R-line of $Mn^{4+}$ ions was observed between 83 and 400 K, at a rate of 0.20 nm $K^{-1}$, while exhibiting low sensitivity to pressure variations[175]. Conversely, the luminescence intensity ratio of this band relative to the phonon progression band demonstrated significant pressure sensitivity, with a maximum of 21% $GPa^{-1}$ at approximately 5.9 GPa. As highlighted above, the sensitivity of transition metal ions to environmental parameters is also influenced by the local symmetry of the crystallographic site they occupy in the host material[90,92,94,95]. This enables the development of materials doped with ions, such as $Cr^{3+}$, situated in crystallographic positions characterized by varying crystal field strengths. These differences lead to pronounced variations in the luminescence behavior under temperature and pressure changes. An illustrative example of this effect can be used $CaAl_{12}O_{19}$:$Cr^{3+}$, which exhibits both narrow-band luminescence associated with strong crystal field sites and broad-band luminescence from $Cr^{3+}$-$Cr^{3+}$ pairs[176]. In this material, the $Cr^{3+}$-$Cr^{3+}$ pairs' band is insensitive to pressure-induced spectral shifts, but it undergoes a thermal shift, allowing for ratiometric pressure sensing with sensitivity as high as 70% $GPa^{-1}$. Simultaneously, the elevation of temperature causes thermalization of the $^4T_2$ level in respect to the $^2E$ level in $Cr^{3+}$ ions at strong crystal field sites, enabling ratiometric temperature measurements with sensitivities up to 1% $K^{-1}$. A similar multi-site emission approach was exemplified by $MgGeO_3$:$Cr^{3+}$, where luminescence arises from two distinct $Ge^{3+}$ crystallographic positions occupied by $Cr^{3+}$ ions[205]. While both sites are characterized by weak crystal fields, variations in $Cr^{3+}$-$O^{2-}$ bond lengths between these sites result in distinguishable $^4T_2 \rightarrow {}^4A_2$ band maxima. Differences in the temperature quenching rates of these emissions, driven by varying activation energies, facilitate ratiometric temperature sensing with sensitivities up to 0.75% $K^{-1}$ at 600 K. Concurrently, pressure-induced modifications in crystal field strength cause spectral shifts in the two bands, enabling ratiometric pressure sensing with a manometric sensitivity of 65% $GPa^{-1}$ at 5 GPa and exceeding 20% $GPa^{-1}$ up to 6 GPa.

In Table 2, we compare different materials used as bifunctional multi-parameter temperature and pressure sensors, including their operating parameters, performance (sensitivity) and working pressure/temperature ranges.

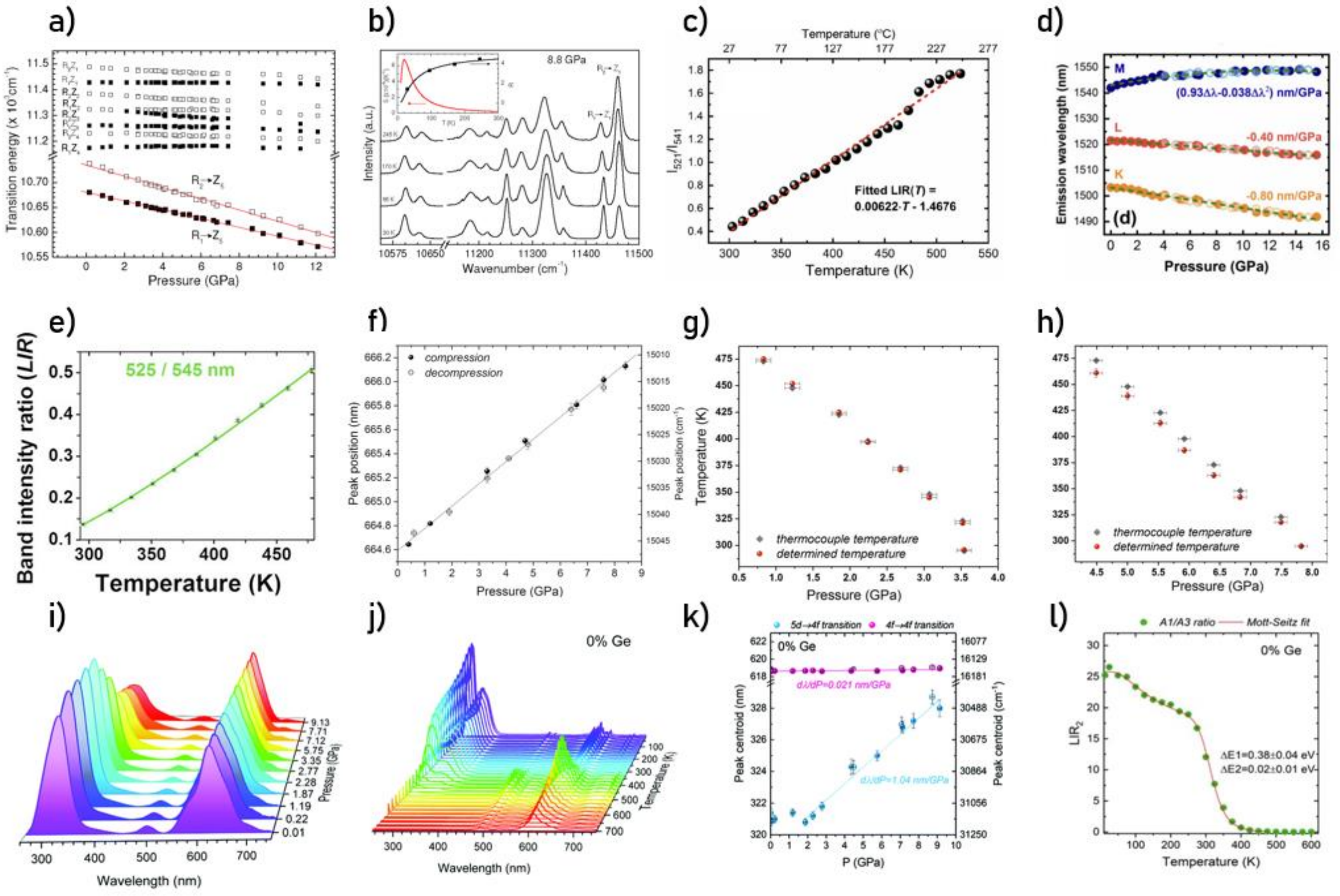

**Figure 5.** Pressure dependence of the energy of the emission lines of $Nd^{3+}$ ions in $Gd_3Sc_2Ga_3O_{12}$:$Cr^{3+}$,$Nd^{3+}$- a)[201]; and corresponding thermal dependence of emission spectra - b)[201]; thermal dependence of LIR ($^2H_{11/2}\rightarrow{}^4I_{15/2}$ and $^4S_{3/2}\rightarrow{}^4I_{15/2}$) of $NaBiF_4$:$Yb^{3+}$,$Er^{3+}$ nanoparticles- c)[200]; and the influence of the pressure on spectral position of the emission bands of this phosphor- d)[200]; thermal dependence of LIR of $Er^{3+}$ ions ($^2H_{11/2}\rightarrow{}^4I_{15/2}$ and $^4S_{3/2}\rightarrow{}^4I_{15/2}$) in $YF_3$:$Yb^{3+}$, $Er^{3+}$ -e) and the influence of pressure on the $Er^{3+}$ bands position -f);

temperature measurements conducted as a function of pressure, using thermocouple mounted in the DAC as a reference and the LIR technique for optical temperature determination; the developed upconverting $YF_3:Yb^{3+}$, $Er^{3+}$ optical sensor was used as an experimental probe of pressure and temperature- g) and h)[199]; emission spectra of $Y_2SiO_5:Pr^{3+}$ measured as a function of pressure -i) and temperature – j)[72]; pressure dependence of the band position of $Pr^{3+}$ ions in $Y_2SiO_5:Pr^{3+}$ - k)[72]; and thermal dependence of LIR of $Pr^{3+}$ emission bands in $Y_2SiO_5:Pr^{3+}$ - l)[72].

**Table 2**. The list of the bifunctional luminescent sensors operating on a different spectroscopic parameters.

| Material | Thermometric parameter | $S_{R,T}$ [% K$^{-1}$] | Manometric parameter | $S_{R,p}$ [% GPa$^{-1}$] | Reference |
|---|---|---|---|---|---|
| $Gd_3Sc_2Ga_3O_{12}$:$Cr^{3+}$, $Nd^3$+ | LIR: $R_{1,2}(^4F_{3/2})\rightarrow Z_1(^4I_{9/2})$ $Nd^{3+}$ | 6.7 % K$^{-1}$ | Bandshift: $R_{1,2}(^4F_{3/2})\rightarrow Z_5(^4I_{9/2})$ $Nd^{3+}$ | 11.3 cm$^{-1}$ GPa$^{-1}$ for $R_1$<br>8.8 cm$^{-1}$ GPa$^{-1}$ for $R_2$; | [201] |
| $LaPO_4$:$Yb^{3+}$, $Tm^{3+}$ | LIR: $^3F_{2,3}\rightarrow{}^3H_6$ $Tm^{3+}$/$^3H_4\rightarrow{}^3H_6$ $Tm^{3+}$ | ~3% K$^{-1}$ | Bandshift: $^3H_4\rightarrow{}^3H_6$ $Tm^{3+}$<br>LIR: $^3H_4\rightarrow{}^3H_6$ $Tm^{3+}$/$^1G_4\rightarrow{}^3H_6$ $Tm^{3+}$ | 0.25 nm GPa$^{-1}$<br><10% GPa$^{-1}$ | [140] |
| $NaBiF_4$:$Yb^{3+}$,$Er^{3+}$ | LIR: $^2H_{11/2}\rightarrow{}^4I_{15/2}$ $Er^{3+}$/$^4S_{3/2}\rightarrow{}^4I_{15/2}$ $Er^{3+}$ | 1.07% K$^{-1}$ | Bandshift: $^4I_{13/2}\rightarrow{}^4I_{15/2}$ $Er^{3+}$<br>LIR: $^2H_{11/2}\rightarrow{}^4I_{15/2}$ $Er^{3+}$/$^4S_{3/2}\rightarrow{}^4I_{15/2}$ $Er^{3+}$ | 0.8 nm GPa$^{-1}$<br>2.09% GPa$^{-1}$ | [200] |
| $YF_3$:$Yb^{3+}$,$Er^{3+}$ micron-sized particles | LIR:<br>$^2H_{11/2}\rightarrow{}^4I_{15/2}$ $Er^{3+}$/$^4F_{9/2}\rightarrow{}^4I_{15/2}$ $Er^{3+}$<br>$^2H_{11/2}\rightarrow{}^4I_{15/2}$ $Er^{3+}$/$^4S_{3/2}\rightarrow{}^4I_{15/2}$ $Er^{3+}$ | <br>1.4% K$^{-1}$<br>1.2% K$^{-1}$ | Bandshift: $^4F_{9/2}\rightarrow{}^4I_{15/2}$ $Er^{3+}$ | 0.2 nm GPa$^{-1}$ | [199] |
| $Y_2Ge_{0.1}Si_{0.9}O_5$: $Pr^{3+}$ | LIR:<br>5d→4f/4f→4f $Pr^{3+}$<br>Lifetime: 5d→4f $Pr^{3+}$ | ~2.5% K$^{-1}$<br>~1.7% K$^{-1}$ | Bandshift: $Pr^{3+}$<br>5d→4f<br>4f→5d<br>4f→4f | <br>1.28 nm GPa$^{-1}$<br>0.43 nm GPa$^{-1}$<br>0.039 nm GPa$^{-1}$ | [72] |
| $Sr_2MgSi_2O_7$:$Dy^{3+}$,$Eu^{2+}$ | FWHM: $4f_6$ $5d_1\rightarrow 4f$ $_7$ $Eu^{2+}$<br>Lifetime: $4f_6$ $5d_1$ $Eu^{2+}$ | 1.29% K$^{-1}$ | Bandshift: $4f_6$ $5d_1\rightarrow 4f$ $_7$ $Eu^{2+}$<br>FWHM: $4f_6$ $5d_1\rightarrow 4f$ $_7$ $Eu^{2+}$ | 8.11 nm GPa$^{-1}$<br>14.8 nm GPa $^{-1}$ | [198] |
| | | | Lifetime: $4f_6$ $5d_1$ $Eu^{2+}$ | 42% $Gpa^{-1}$ | |
| $Cs_2Ag_{0.6}Na_{0.4}InCl_6$ $Bi^{3+}$ | FWHM: exciton emission band | 0.2547 nm $K^{-1}$ | Bandshift: exciton emission band<br>FWHM: exciton emission band | 112 nm $GPa^{-1}$<br>31.5 nm $GPa^{-1}$ | [197] |
| $Gd_2ZnTiO_6$:$Mn^{4+}$ | FWHM: $^2E \rightarrow {}^4A_2$ $Mn^{4+}$<br>Lifetime: $^2E$ $Mn^{4+}$ | 0.34% $K^{-1}$<br>2.43% $K^{-1}$ | Bandshift: $^2E \rightarrow {}^4A_2$ $Mn^{4+}$ | 1.11 nm $GPa^{-1}$ | [25] |
| $YPO_4$:$Yb^{3+}$,$Er^{3+}$ | LIR: $^2H_{11/2} \rightarrow {}^4I_{15/2}$ $Er^{3+}$/$^4S_{3/2} \rightarrow {}^4I_{15/2}$ $Er^{3+}$ | 1.25% $K^{-1}$ | Bandshift: $^4I_{13/2} \rightarrow {}^4I_{15/2}$ $Er^{3+}$ | 0.539 nm $GPa^{-1}$ | [196] |
| $GdTaO_4$: $Nd^{3+}$ | LIR: $R_{1,2}(^4F_{3/2}) \rightarrow Z_5(^4I_{9/2})$ $Nd^{3+}$ | 0.148% $K^{-1}$ | Bandshift: $R_2(^4F_{3/2}) \rightarrow Z_5(^4I_{9/2})$ $Nd^{3+}$ | 1.34 nm $GPa^{-1}$ | [195] |
| $Y_2O_3$:$Yb^{3+}$-$Nd^{3+}$ &<br>$Y_2O_3$:$Yb^{3+}$-$Er^{3+}$ | LIR $Nd^{3+}$ 759/872 nm<br>LIR $Er^{3+}$ 661/563 nm | | LIR $Er^{3+}$ 657/547 nm | | [194] |
| $La_3Mg_2SbO_9$:$Mn^{4+}$ | Lifetime: $^2E$ $Mn^{4+}$ | 2.52 % $K^{-1}$ | Bandshift: $^2E \rightarrow {}^4A_2$ $Mn^{4+}$ | 1.2 nm $GPa^{-1}$ | [193] |
| $CaAl_{12}O_{19}$:$Cr^{3+}$ | LIR: $^4T_2 \rightarrow {}^4A_2$ $Cr^{3+}$/$^4T_2$, $A_2 \rightarrow {}^4A_2$, $A_2$ $Cr^{3+}$ | ~1.1% $K^{-1}$ | LIR: $^4T_2 \rightarrow {}^4A_2$ $Cr^{3+}$/$^2E \rightarrow {}^4A_2$ $Cr^{3+}$<br>Lifetime: $^2E$ $Cr^{3+}$ | 70% $GPa^{-1}$<br>16% $GPa^{-1}$ | [176] |
| $K_2Ge_4O_9$:%$Mn^{4+}$ | Bandshift:R-line $^2E \rightarrow {}^4A_2$ $Mn^{4+}$ | 0.02 nm $K^{-1}$ | LIR $^2E \rightarrow {}^4A_2$ $Mn^{4+}$/$^2E \rightarrow {}^4A_2$ $Mn^{4+}$<br>Lifetime: $^2E$ $Mn^{4+}$ | 21.7 % $GPa^{-1}$<br>12 % $GPa^{-1}$ | [175] |

## 5. Strategies for development of bifunctional optical sensors

As discussed above, the scientific literature contains numerous reports on bifunctional luminescence thermometers-manometers. However, many of these studies appear to reflect outcomes of trial-and-error methodologies rather than systematic, goal-oriented design strategies. To develop materials with effective bifunctional sensing performance, it is essential to adopt a more targeted approach based on a comprehensive understanding of the underlying mechanisms in the materials and the specific requirements and constraints of the intended applications. The most important challenge is to perform pressure-temperature cross-dependence measurements, to truly reveal their bilateral (in)dependences[199].

Previous studies have demonstrated that luminescent materials are not ideal for universal sensors capable of operating across a broad range of pressures and temperatures. Instead, application-specific solutions are preferred, allowing for the optimization of these materials for specific conditions. This limitation arises from the inherent characteristics of the sensing methodologies. Whether based on luminescence intensity ratios, luminescence kinetics, or other parameters, certain conditions can lead to the complete extinction of luminescence intensity or deactivation of excited levels, limiting the sensor's range. Furthermore, the sensitivity of a sensor is directly proportional to the rate of change of its luminescence properties under the applied stimuli. Consequently, sensor designers face a trade-off: creating sensors with a broad operating range but low sensitivity, or maximizing sensitivity at the expense of range.

In many practical applications, the second approach is more favorable, as the requirement for a wide operating range is often unnecessary. For instance, when pressure changes are expected to be minimal, the LIR parameter based on thermally-coupled levels can serve as an effective determinant of temperature. Conversely, in systems where temperature is stabilized or exhibits minimal fluctuation, the spectral position of bands, as used in ruby-based pressure determination in diamond anvil cells, can be employed for precise pressure sensing.

A more universal solution involves the use of different parameters to independently determine pressure and temperature. This can be achieved by exploiting physical mechanisms that ensure the insensitivity of one spectroscopic parameter to a specific physical stimulus, leading to materials with selective sensitivity. In the case of the lanthanide-activated sensor materials, good example of almost ideal bi-functional *p-T* sensors based on two, nearly-independent parameters are materials based on $Er^{3+}$ luminescence (both down-shifting and up-conversion), such as the discussed $YF_3$:$Yb^{3+}$,$Er^{3+}$[199]. This is because, in such cases the LIR of $Er^{3+}$ ($^2H_{11/2}$/$^4S_{3/2}$ → $^4I_{15/2}$) is very sensitive to temperature and nearly insensitive to pressure. Whereas, the spectral position of $Er^{3+}$ emission bands in barely affected by temperature, but quite significantly dependent on pressure. The validity of such findings have been confirmed experimentally by *p-T* cross-calibration experiments under extreme conditions of pressure and temperature, where both state functions could be successfully and independently monitored[199]. A notable example for the sensors based on the d-block metal ions is the ratiometric pressure sensing in $Cr^{3+}$-doped materials with weak crystal fields[26,78,206]. In these systems, pressure-induced shortening of $Cr^{3+}$-$O^{2-}$ bond lengths increases crystal field strength, causing a spectral shift in the $^4T_2$ →$^4A_2$ band[26]. This shift also results in a monotonic change in the intensity ratio of signals recorded in two spectral windows. Importantly, temperature changes are not expected to affect the crystal field strength significantly, ensuring that the LIR parameters remain insensitive to temperature variations. This approach has been experimentally validated, with thermal insensitivity characterized using the TIMF (Thermal Insensitivity Manometric Factor) parameter. The TIMF indicates the temperature change required to induce spectroscopic changes equivalent to those caused by a 1 GPa of pressure change. Its inverse serves as an analogous metric for pressure sensitivity. On the other hand, the PITF (pressure invariable thermometeric factor) parameter could quantitatively characterize the performance, and pressure-independence of the bi-functional temperature sensors working under extreme conditions of pressure. The combination of $Cr^{3+}$ based pressure sensor with the ratiometric temperature sensor like $Er^{3+}$ based can be a promising direction for further research.

Another promising research direction involves using the luminescence kinetics of transition metal ions, particularly $Mn^{4+}$ ions, to achieve selective sensitivity[173,175][. The high activation energy for nonradiative depopulation of the excited $^2E$ level in $Mn^{4+}$ can be controlled through host material composition, potentially enabling thermal stability of the $^2E$ lifetime across a wide temperature range[27,172,173]. Simultaneously, the luminescence kinetics of the $^2E$ level in $Mn^{4+}$ (and other $3d^3$ ions) exhibit favorable pressure sensitivity due to spin-orbit coupling between the $^2E$ and $^4T_2$ levels. The $^2E \rightarrow {}^4A_2$ transition is spin-forbidden, whereas $^4T_2 \rightarrow {}^4A_2$ is spin-allowed. Increased crystal field strength reduces the coupling, leading to a pressure-induced increase in the $^2E$ lifetime. This phenomenon allows the development of kinetic-based luminescence sensors of pressure with minimal temperature sensitivity. Despite its potential, this approach remains underexplored and warrants further investigation.

## 6. Conclusions

Inorganic bifunctional contactless temperature and pressure sensors are promising materials for simultaneous, precise and rapid detection of two physical parameter changes in extreme, inaccessible systems. The interest in developing new materials for simultaneous pressure and temperature sensing is still growing due to the potential applications in aerospace, energy and biomedical sectors. We have gathered and presented up-to-date results in bifunctional optical sensing of pressure and temperature based on the investigation of single or multi-parameter approaches.

## Acknowledgements

This work was supported by the Foundation for Polish Science under First Team FENG.02.02-IP.05-0018/23 project with funds from the 2nd Priority of the Program European Funds for Modern Economy 2021-2027 (FENG), and Polish National Science Center (grant no. 2023/50/E/ST5/00021 and 2023/51/D/ST5/00579).